\documentclass[aps,prl,preprint,endfloats,superscriptaddress]{revtex4}

\usepackage{graphicx}

\begin{document}

\title{Magnetic trapping and Zeeman relaxation of imidogen (NH $X^3\Sigma^{-}$)}

\author{Wesley C. Campbell}
\email[]{wes@cua.harvard.edu}
\author{Edem Tsikata}
\author{Laurens van Buuren}
\altaffiliation{Current address: Max-Planck-Institut f\"ur
  Quantenoptik, Garching, Germany}

\affiliation{Department of Physics, Harvard University, Cambridge,
  Massachusetts 02138, USA}
\affiliation{Harvard-MIT Center for Ultracold Atoms, Cambridge,
  Massachusetts 02138, USA}
\author{Hsin-I Lu}
\affiliation{Division of Engineering and Applied Sciences, Harvard
  University, Cambridge, MA 02138, USA}
\affiliation{Harvard-MIT Center for Ultracold Atoms, Cambridge,
  Massachusetts 02138, USA}
\author{John M. Doyle}
\affiliation{Harvard-MIT Center for Ultracold Atoms, Cambridge,
  Massachusetts 02138, USA}
\affiliation{Department of Physics, Harvard University, Cambridge,
  Massachusetts 02138, USA}

\date{\today}

\begin{abstract}

Imidogen (NH) radicals are magnetically trapped and their Zeeman
relaxation and energy transport collision cross sections with helium are
measured.  Continuous buffer-gas loading of the trap is direct from a
room-temperature molecular beam.  The Zeeman relaxation (inelastic)
cross section of magnetically trapped electronic, vibrational and
rotational ground state imidogen in collisions with $\mathrm{^{3}He}$
is measured to be $3.8\pm1.1\times10^{-19}\mbox{ cm}^2$ at 710 mK.
The NH-He energy transport cross section is also measured, indicating
a ratio of diffusive to inelastic cross sections of $\gamma =
7\times10^{4}$, in agreement with recent theory \cite{KremsPRA03}.

\end{abstract}


\maketitle

Driven by the promise of new physics and applications, the field of
cold molecular physics has undergone tremendous growth in the past
decade \cite{KremsEPJD04,MeijerIRPC03}.  In particular, polar
molecules have been proposed \cite{DeMillePRL02,ZollerNATPHYS06} as
qubits for quantum information processing and for studies of highly
correlated condensed matter systems \cite{ZollerNATPHYS06b}.  Cold
molecules are also very promising candidates for fundamental physics
measurements, such as the search for the electric dipole moment of the
electron \cite{LabzowskyJPB95} and time variation of the
electron-proton mass ratio \cite{KorobovPRA05}.  Chemistry with cold
molecules may be possible to observe in a new quantum regime where the
large deBroglie wavelength and long interaction times of reactants can
reopen chemical reaction pathways through tunneling
\cite{DalgarnoCPL01,KremsIRPC05}.  A key to realizing these new
phenomena is the production of cold, trapped, high-density samples of
polar molecules.

The complexity inherent in molecules, compared to atoms, makes working
with them difficult; the rich internal structure of molecules opens
decay modes that, for example, could make evaporative cooling
difficult to achieve \cite{BohnPRL02}.  Preparation of cold, dense
samples of trapped molecules should provide a pathway to measure cold
molecule collisions, critical to elucidating the suitability of
molecules for evaporative \cite{BohnPRA06,GroenenboomJCP05,KajitaPRA06} and sympathetic
cooling \cite{HutsonPRL06}.  So far, several methods have been used to
create (ultra-)cold molecules. Production of ultracold molecules from
ultracold atoms has been realized through photo and Feshbach
association \cite{DeMillePRL05,WiemanNATURE02}.  Direct cooling
or slowing
of initially hot polar molecules has also been demonstrated using Stark
deceleration \cite{MeijerPRL99,MeijerIRPC03}, optical slowing
\cite{BarkerJPB06}, billiard-like collisions \cite{ChandlerSCIENCE03},
and buffer-gas cooling \cite{DoyleNATURE98}.  Buffer gas cooling uses
cryogenic helium gas to cool hot molecules (or atoms). When done in
the presence of a magnetic trapping field this can cause the molecules
to fall into the conservative trap potential.  The first successful
trapping of polar molecules was accomplished in our group with
buffer-gas loading of CaH, and we know of several other laboratories
using buffer-gas loading of magnetic traps
\cite{PetersJPB06,deCarvalhoCJP05,CesarBJP01}.  Other molecules (VO
\cite{DoyleJCP98}, CaF \cite{DoylePRL05} and CrH
\cite{PetersUnpublished}) have also been studied with buffer-gas
loading but in all those cases there were loss mechanisms that
prevented trapping for extended periods of time.

Despite great progress in the field of cold molecules, no technique
has yet realized trapped densities sufficiently high to observe polar
molecule-molecule collisions. However, the buffer gas method allowed
studies of cold atom-polar molecule spin relaxation of trapped
molecules (He-CaH and CaF \cite{DoyleNATURE98,DoylePRL05}).  These
measurements in combination with theory began to uncover the
fundamental processes of cold molecule collisions in traps. CaH has a
$^2\Sigma$ ground state, perhaps the simplest type of magnetically
trappable molecule. Although this provided important information on
the nature of cold molecule collisions, it was incomplete as there are
a variety of molecules.  For example, $^3\Sigma$ molecules carry with
them new internal dynamics, such as the spin-spin interaction, which
allows for direct coupling of the rotation during a collision (unlike
the $^2\Sigma$ case) \cite{DalgarnoJCP04}.  The current vigorous
pursuit of high density samples of cold molecules has led several
groups to the imidogen (NH) radical.  Cold, trapped imidogen is being
studied theoretically \cite{KremsPRA03,KremsJCP05,HutsonPRL06,KajitaPRA06} and
pursued experimentally
\cite{MeijerJPB06,DoyleEPJD04,LewandowskiHeraeus}, and a scheme for
continuous loading of imidogen into a magnetic trap has been proposed
for Stark deceleration \cite{MeijerPRA03}.

In this Letter, we demonstrate magnetic trapping of ground state
($X^3\Sigma^{-}$) imidogen radicals and make a direct measurement of
the spin relaxation rate in collisions with $^3$He.  Imidogen is
continuously loaded directly from a room-temperature molecular beam
into a magnetic trap \emph{via} buffer-gas cooling.  More than
$10^{\mathrm{8}}$ molecules are loaded into the trap and are observed
for longer than 1 s, with $1/e$ lifetimes exceeding 200 ms.  The
energy transport collision rate is also measured, allowing the
determination of the ratio of the diffusive to inelastic cross
sections in this system, found to be $\gamma =
\sigma_{\mathrm{d}}/\sigma_{\mathrm{in}} = 7\times10^{4}$.

We chose to study imidogen (NH) due to its internal structure,
predicted collisional properties, and for technical reasons.  The
$X^{3}\Sigma^{-}$ ground state has a $2\mu_{\mathrm{B}}$ magnetic
moment and 1.38 Debye electric dipole moment.  The imidogen radicals
can be produced in an ammonia discharge \cite{UbachsCJP84} and
detected in absorption or fluorescence on the $A^{3}\Pi
\leftrightarrow X^{3}\Sigma^{-}$ transition (336 nm, $1/A=400$ ns).
It has been predicted \cite{KremsPRA03} that $^{3}\Sigma$ molecules
with relatively large rotational splitting and weak spin-spin
interaction (such as the imidogen radical) will be less likely to
undergo collision induced Zeeman transitions.  Another important
feature of the NH system is that imidogen can be produced in a high
flux beam with a room temperature discharge source \cite{DoyleEPJD04}.
This provides a new challenge - the introduction of this beam into our
very low temperature trapping region.  This was a key experimental
challenge that was met with success and can be applied to numerous
other species.

\begin{figure}
\includegraphics[width=6in]{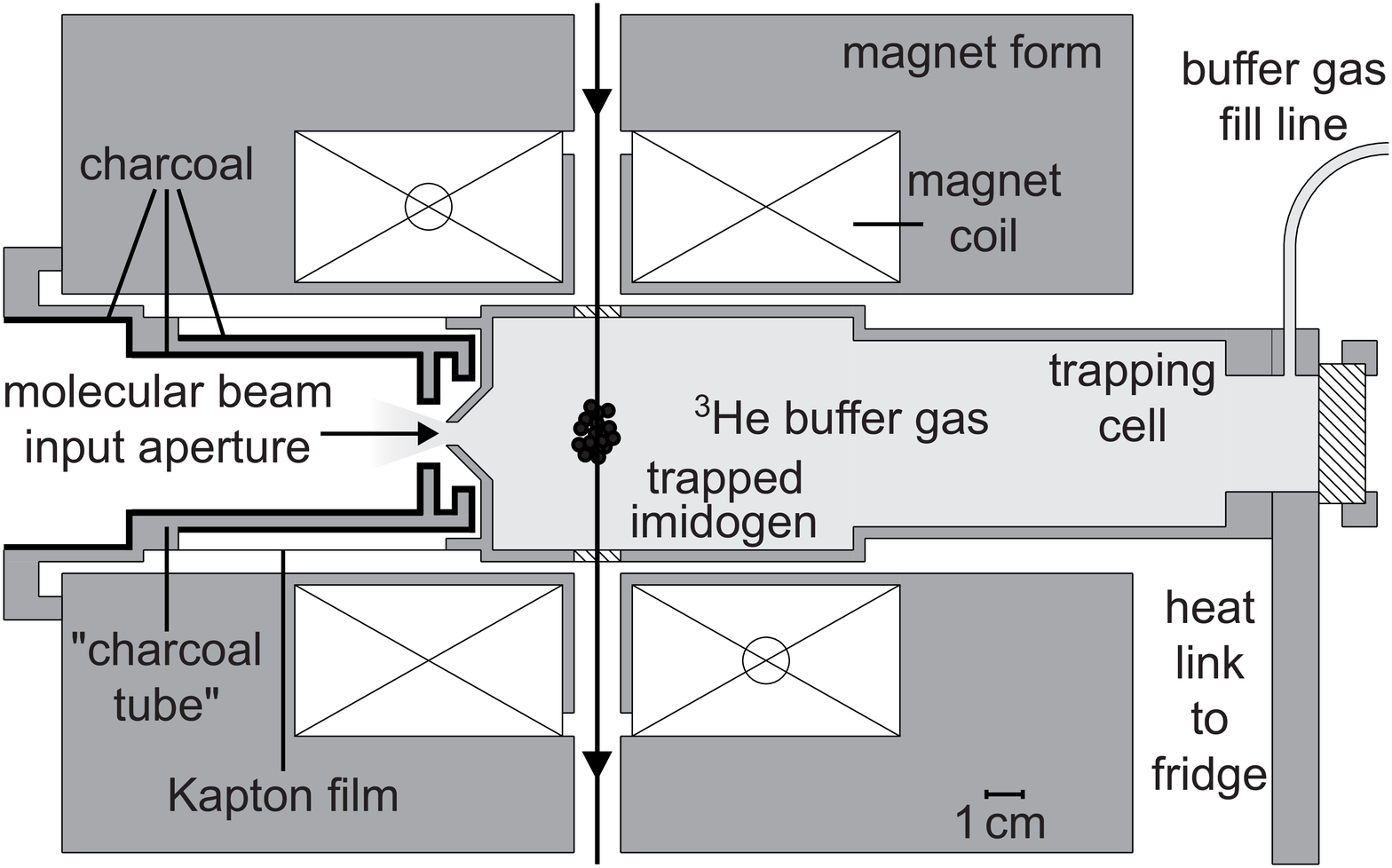}
\caption{Schematic diagram of the magnetic trap, buffer-gas cell and
  charcoal tube in vacuum.  The imidogen beam is produced by flowing
  ammonia from a pulsed valve through a slit glow discharge 12 cm to
  the left of the molecular beam input aperture.  The radical beam
  propagates through openings in the 77 K and 4 K blackbody radiation
  shields (not shown), through a section of the magnet bore, through
  the molecular beam input aperture and into the trapping cell.  The
  trap magnet surrounds the trapping cell and is described in
  Ref. \cite{DoyleRSI04}.\label{AppDiagram}}
\end{figure}

Our apparatus is centered around a cold cell made from copper
\footnote{Alloy C10100, annealed in forming gas.} and is thermally
disconnected from the 4 K magnet surrounding it (see
Fig. \ref{AppDiagram}).  The maximum trap depth available is 3.9
Tesla.  Windows at the magnet midplane allow optical access for laser
beams and a fluorescence collection lens.  The cell is thermally
connected to a $^3$He refrigerator, giving the cell a base temperature
of 450 mK.

Buffer gas enters the cell through a fill line and exits out a 3 mm
diameter opening in the side of the cell that faces the discharge
source (the ``molecular beam input aperture" shown in
Fig. \ref{AppDiagram}).  Helium is continuously supplied to the cell
in order to maintain a constant helium atom density in the trapping
region.

The constant flow of helium buffer gas out of the cell aperture poses
a significant technical problem.  The helium gas can scatter imidogen
radicals out of the incident NH beam before they enter the trapping
cell.  In order to maintain sufficient vacuum just outside the cell, a
charcoal coated copper tube (``charcoal tube") is used to pump away
the escaping helium.  The charcoal tube is held at 4 K so as to act as
a low-profile high-speed vacuum pump.  This eliminates any significant
scattering of imidogen radicals.

The helium buffer gas density in the cell is determined by the rate at
which we flow helium into the cell and the conductance out of the
molecular beam input aperture.  The conductance of the molecular beam
input aperture was measured using a fast ion gauge, agreeing with our
calculated theoretical value.  This system allows both experimental
control and absolute knowledge of the buffer-gas density to better
than 20\%.

The imidogen radicals in the trapping region are detected using laser
induced fluorescence (LIF) excited on the
$\left|A^{3}\Pi_{2},v^{\prime}=0,
N^{\prime}=1\rangle\right. \leftarrow
\left|X^{3}\Sigma^{-},v^{\prime\prime}=0,N^{\prime\prime}=0\rangle\right.$
transition.  The light source is a CW dye laser frequency doubled in
BBO in an external buildup cavity.  The excitation laser beam enters
the cell and passes through the trap center along a diameter before
exiting.  Fluorescence from $A \rightarrow X$ is collected by a
midplane lens perpendicular to both the laser and molecular beam axes
and imaged to the face of a photomultiplier tube (PMT).  In this way,
time-resolved narrow-band spectra of trapped imidogen radicals are
gathered.  As the spherical quadrupole field of the trap magnet is
strongly inhomogeneous and the LIF line frequency is field-dependent,
laser frequency maps directly to distance from the trap center.

\begin{figure}
\includegraphics[width=6in]{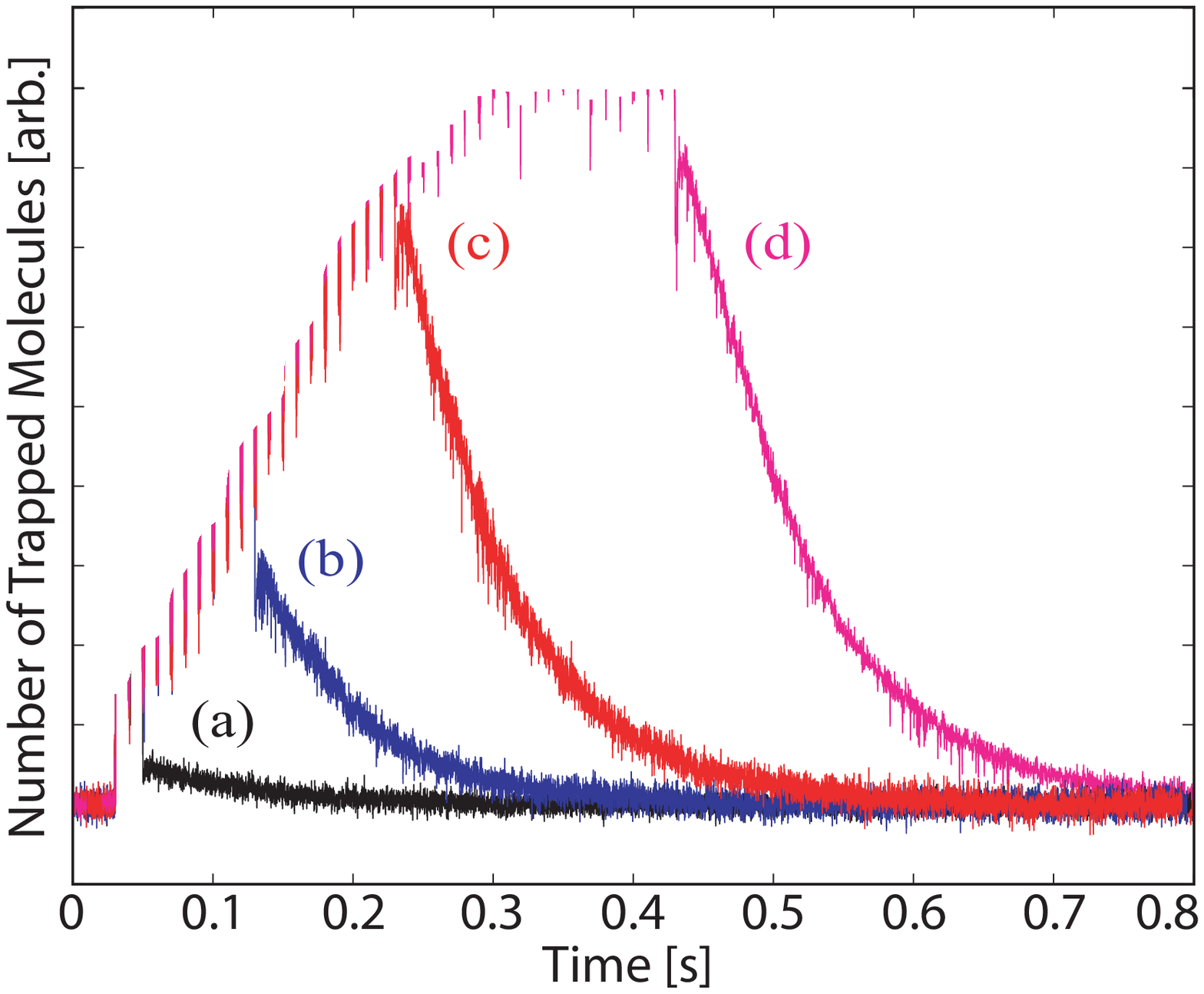}
\caption{Time profiles of the fluorescence signal gathered by the PMT.
  The curves correspond to different durations of the molecular beam
  loading pulse of (a) 10 ms, (b) 100 ms, (c) 200 ms and (d) 400 ms.
  The dissociation discharge causes periodic overranging of the
  signal during loading.\label{Loading}}
\end{figure}

Figure \ref{Loading} shows a series of time profiles of the number of
trapped imidogen molecules for several different loading times.  This
signal is simply the PMT count rate versus time using an LIF
excitation frequency set to the peak of the trapped imidogen spectral
feature.  As this is a fluorescence experiment, the absolute number of
molecules is difficult to determine accurately.  By calculating the
efficiency of our detection system, excitation probability, and
detected molecule fraction we can, however, put a lower bound on our
initial number of trapped imidogen radicals of $> 10^8$.  The data
shown in Fig. \ref{Loading} was taken with $^4$He, so the temperature
of the buffer-gas cell was elevated to 730 mK (in order to maintain
sufficient $^4$He density) at a trap depth of 3.3 Tesla.  Each trace
corresponds to a different duration of loading from the molecular beam
source - the tail ends of all four traces fit well to a
single-exponential decay with time constant of $90\pm10$ ms.  For
loading times less than the molecular trap lifetime, the number of
trapped molecules increases nearly in proportion with with loading
time (see curves (a), (b) and (c)).  This demonstrates one of the key
features of the continuous molecular beam loading technique: as the
lifetime of molecules in the trap increases, the loading time can be
increased resulting in more trapped molecules.  Ablation sources for
buffer-gas loading experiments have to this point shown insignificant
increases in numbers trapped from multiple loading pulses, likely due
to the violent nature of the ablation plume in the trapping region.
Fig. \ref{Loading}(c,d) shows that as the loading time exceeds the
trap lifetime the signal height saturates and there is no longer any
significant increase in the number of trapped molecules, as expected.

Trapping is spectroscopically verified by tuning the LIF laser to be
resonant with high-field-seeking (HFS) molecules and comparing time
profiles to low-field seekers (LFS).  The HFS molecules quickly leave
the trap and are undetectable after 10 ms while the lifetimes of the
LFS molecules are enhanced to more than 200 ms at our lowest loading
temperature.  The spatial sensitivity of our fluorescence collection
precludes fitting the spectra for temperature, but we have previously
demonstrated rotational and translational thermalization of imidogen
to the buffer gas temperature \cite{DoyleEPJD04}.  Finally, as
described below, only trapped molecules produce the lifetimes vs. trap
depths that we observe.

\begin{figure}
\includegraphics[width=6in]{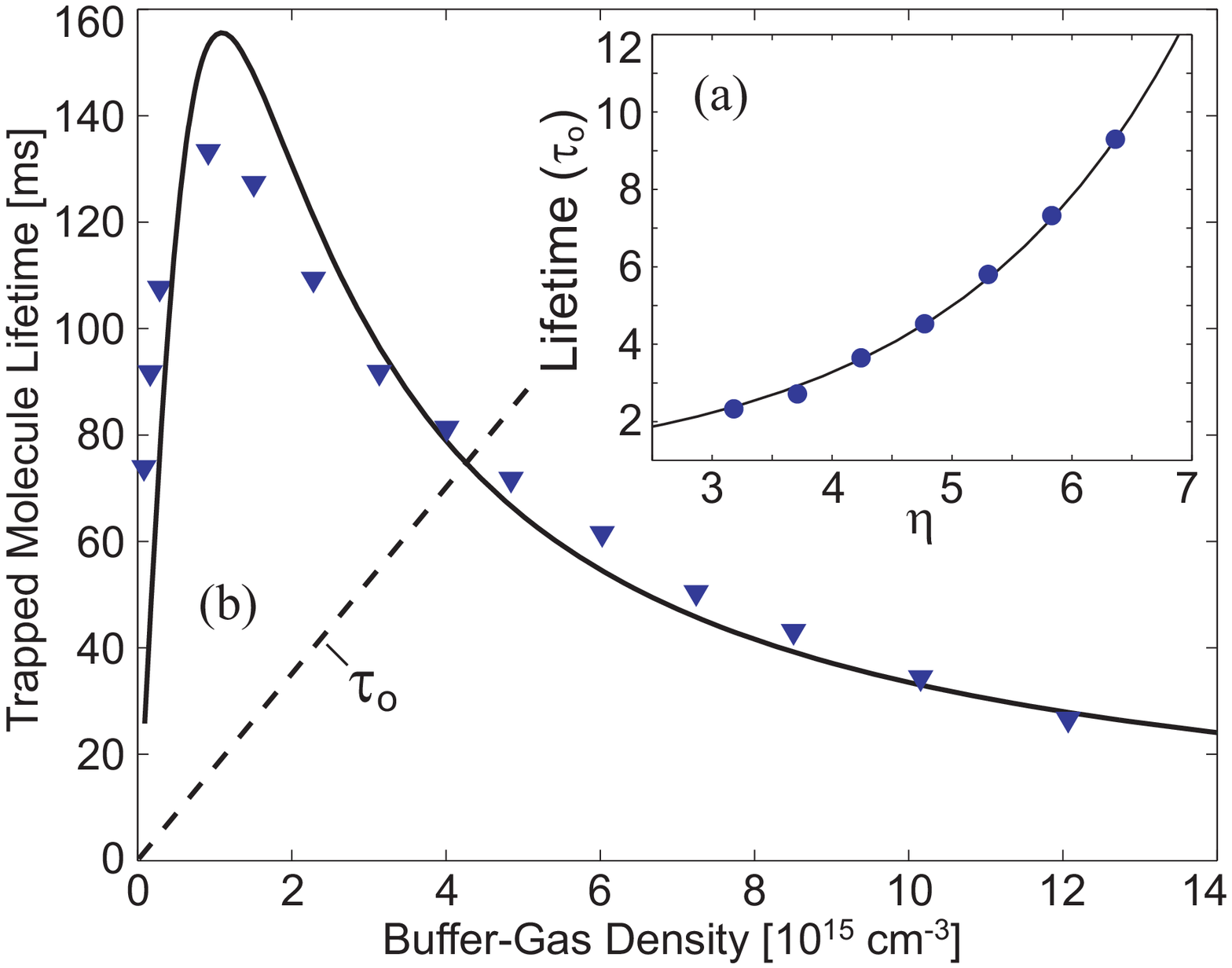}
\caption{Lifetimes for magnetically trapped imidogen radicals vs. (a)
  trap depth and (b) buffer-gas density. Trapped molecule lifetimes
  were obtained by fitting the fluorescence to a single-exponential
  decay.  $\tau_o$ is the field-free diffusion lifetime calculated from
  Eq. \ref{tauo}.  The solid curve in (b) is a two-parameter fit of the form
  $1/\tau_{\mathrm{eff}} = A/n + nk_{\mathrm{in}}$ and yields the
  inelastic cross section.\label{tauvsn}}
\end{figure}

In the absence of collisional losses, the trap lifetime will increase
significantly as trap depth increases and the magnetic trapping field
holds molecules in the center of the cell.  With background buffer gas
present, there is an additional lifetime lengthening factor as
collisions with helium atoms slow motion of trapped molecules to the
walls, essentially enforcing diffusive motion.  Fig. \ref{tauvsn}(a)
(inset) shows the measured trap lifetime as a function of $\eta \equiv
\mu B_{\mathrm{max}} / k_{\mathrm{B}} T$, where $\mu B_{\mathrm{max}}$
is the trap depth.  The lifetime is plotted in units of the field-free
diffusion lifetime in the cell, $\tau_o$, given by \cite{Hasted}
\begin{equation}
\tau_o = \frac{16n\sigma_{\mathrm{d}}}{3\sqrt{2\pi}} \sqrt{\frac{m_{\mathrm{red}}}{k_{\mathrm{B}} T}}
\left[ \left( \frac{\alpha_1}{R}\right)^2 + \left( \frac{\pi}{h}\right)^2 \right]^{-1}
\label{tauo}
\end{equation}
where $n$ is the $^3$He buffer-gas density, $\sigma_{\mathrm{d}}$ is
the thermal average of the diffusion cross section, $R$ and $h$ are
the internal radius and length of the cell, $m_{\mathrm{red}}$ is the
reduced mass of the NH-$^{3}$He system and $\alpha_1$ is the first
root of the Bessel function $J_0$ of order zero.  For the data in
Fig. \ref{tauvsn}(a) the buffer gas density was $8.5 \times 10^{14}$
cm$^{-3}$ and the temperature was 690 mK.  The solid curve is a fit of
a numerical solution to the diffusion equation including the trap
potential \cite{JonathanThesis} and the only fitting parameter is the
cross section, yielding a measurement of $\sigma_{\mathrm{d}} =
2.7\pm0.8 \times 10^{-14}$ cm$^2$.  The quoted uncertainty is
systematic and dominated by uncertainty in the buffer-gas density.
Multiplying this by the average relative velocity of NH-$^3$He gives
the rate coefficient for energy transport, $k_{\mathrm{d}} = 2.1 \pm
0.6 \times 10^{-10}\mbox{ cm}^3\mbox{s}^{-1}$.  This is in good
agreement with the 0.5 K energy transport collision rate of
$1.49\times10^{-10}\mbox{ cm}^3\mbox{s}^{-1}$ calculated by Krems
\emph{et al.} in Ref. \cite{KremsPRA03}.

Figure \ref{tauvsn}(b) shows the measured trap lifetime for different
buffer-gas densities.  The peak lifetime in this study is somewhat low
since cell was heated to 710 mK in order to maintain a constant
temperature throughout the full buffer-gas density range (a technical
artifact of the helium gas supply system).  Trapped imidogen lifetimes
for the \emph{base} temperature of the cell exceed 200 ms.  As the
density of the $^3$He buffer gas is increased from $10^{14}$ to around
$10^{15}$ cm$^{-3}$ the lifetime of the trapped molecules increases
due to the diffusion effect mentioned above.  However, as the $^3$He
density is increased past $10^{15}$ cm$^{-3}$ it is seen that the
lifetime decreases.  This decrease is due to collision-induced Zeeman
relaxation of the imidogen radicals as they collide with $^3$He.  The
time profiles still show good single-exponential decay behavior for
these higher buffer-gas densities.

To extract a rate constant for these inelastic collisions, we fit a
model curve to the data in Fig. \ref{tauvsn}.  After being loaded, the
time profile of the molecule number in the trap can be modeled as
$N(t) = N_oe^{-tA/n}e^{-tnk_{\mathrm{in}}}$ where $n$ is the buffer
gas density, $N_o$ is the initial number of imidogen radicals after
loading, and $A$ and $k_{\mathrm{in}}$ are fitting parameters
corresponding to diffusion enhancement of the lifetime and inelastic
collision loss, respectively.  This can be rewritten as a single
exponential with a time constant given by $1/\tau_{\mathrm{eff}} = A/n
+ nk_{\mathrm{in}}$, which gives the form for the curve fitted to the
data in Fig. \ref{tauvsn}.  The constant $k_{\mathrm{in}}$ is the
collision induced Zeeman relaxation rate at 710 mK and is found to be
$k_{\mathrm{in}}=3.0\pm0.9\times10^{-15}\mbox{ cm}^3\mbox{s}^{-1}$,
corresponding to an inelastic cross section of
$\sigma_{\mathrm{in}}=3.8\pm1.1\times10^{-19}\mbox{ cm}^2$ and
therefore a ratio of diffusive to inelastic cross sections of $\gamma
= 7\times10^4$.  The uncertainties are systematic and are dominated by
uncertainty in the absolute buffer-gas density.  This inelastic
collision rate is an order of magnitude higher than the 500 mK value
of $4.20\times10^{-16}\mbox{ cm}^3\mbox{s}^{-1}$ for 100 Gauss given
by Krems \emph{et al.}  \cite{KremsPRA03} and may be larger due to the
strong temperature dependence of the predicted shape resonance.
Furthermore, the rate we measure is averaged over the magnetic field
range of our trap, and inelastic collision cross sections are predicted
to become strongly field-dependent just below this collision energy
\cite{KremsPRA03,KremsJCP05}.

In summary, we have demonstrated magnetic trapping of ground state
imidogen molecules using buffer gas loading from a molecular beam.
The energy transport (diffusion) and inelastic collision rate
constants for NH-$^3$He have been measured

resulting in a ratio of elastic to inelastic collision rates of
$7\times10^4$, which indicates that imidogen radicals should be
amenable to the pumping out of the buffer gas and thermal isolation.
It is particularly interesting to note that we are able to trap
imidogen near the peak of the narrow shape resonance predicted by
Krems \emph{et al.} in Ref. \cite{KremsPRA03}.  A factor of two
decrease (or increase) in collision energy would decrease the
inelastic rate by more than a factor of ten, resulting in a increase
in $\gamma$ to nearly $10^{6}$.  This bodes very well for both
increasing the number of trapped molecules and perhaps for sympathetic
cooling of large numbers of imidogen radicals and other hydrides to
the ultracold regime.

\begin{acknowledgments}
We thank Katsunari Enomoto and Michael Gottselig for their valuable
assistance in the design and fabrication of the apparatus and also
greatly acknowledge Colin Connolly for his construction of the $^3$He
refrigerator used in these experiments.  The authors are grateful to
B. Friedrich and R. Krems for discussions and careful reading of the
manuscript and G. C. Groenenboom for helpful discussions.  This work
was supported by the U.S. Department of Energy under Contract
No. DE-FG02-02ER15316 and the U.S. Army Research Office.
\end{acknowledgments}

\bibliography{Wesbib.bib}

\end{document}